\documentclass[prb,twocolumn,showpacs,preprintnumbers,amsmath,amssymb]{revtex4}

\usepackage{graphicx}

\begin{document}
\title{Stability of extended defects on boron nitride and graphene
monolayers: the role of chemical environment} 

\author{L. C. Gomes}

\author{S. S. Alexandre}

\author{H. Chacham}

\author{R. W. Nunes}\thanks{corresponding author}
\email{rwnunes@fisica.ufmg.br}

\affiliation{Departamento de F\'{\i}sica,  Universidade Federal de Minas
Gerais, CP 702, 30123-970, Belo Horizonte, MG, Brazil}

\date{\today}

\begin{abstract}
We perform {\it ab initio} calculations that indicate that the
relative stability of antiphase boundaries (APB) with armchair and
zigzag chiralities in monolayer boron nitride (BN) is determined by
the chemical potentials of the boron and nitrogen species in the
synthesis process. In an N-rich environment, a zigzag APB with N-rich
core is the most stable structure, while under B-rich or intrinsic
growth conditions, an armchair APB with stoichiometric core is the
most stable. This stability transition is shown to arise from a
competition between homopolar-bond (B-B and N-N) and elastic energy
costs in the core of the APBs. Moreover, in the presence of a carbon
source we find that a carbon-doped zigzag APB becomes the most stable
boundary near the N-rich limit. The electronic structure of the two
types of APBs in BN is shown to be particularly distinct, with the
zigzag APB depicting defect-like deep electronic bands in the band
gap, while the armchair APB shows bulk-like shallow electronic bands.
\end{abstract}

\pacs {73.22.-f, 73.20.Hb, 71.55.-i}

\maketitle

\begin{figure}
\centering \includegraphics[width=8.0cm]{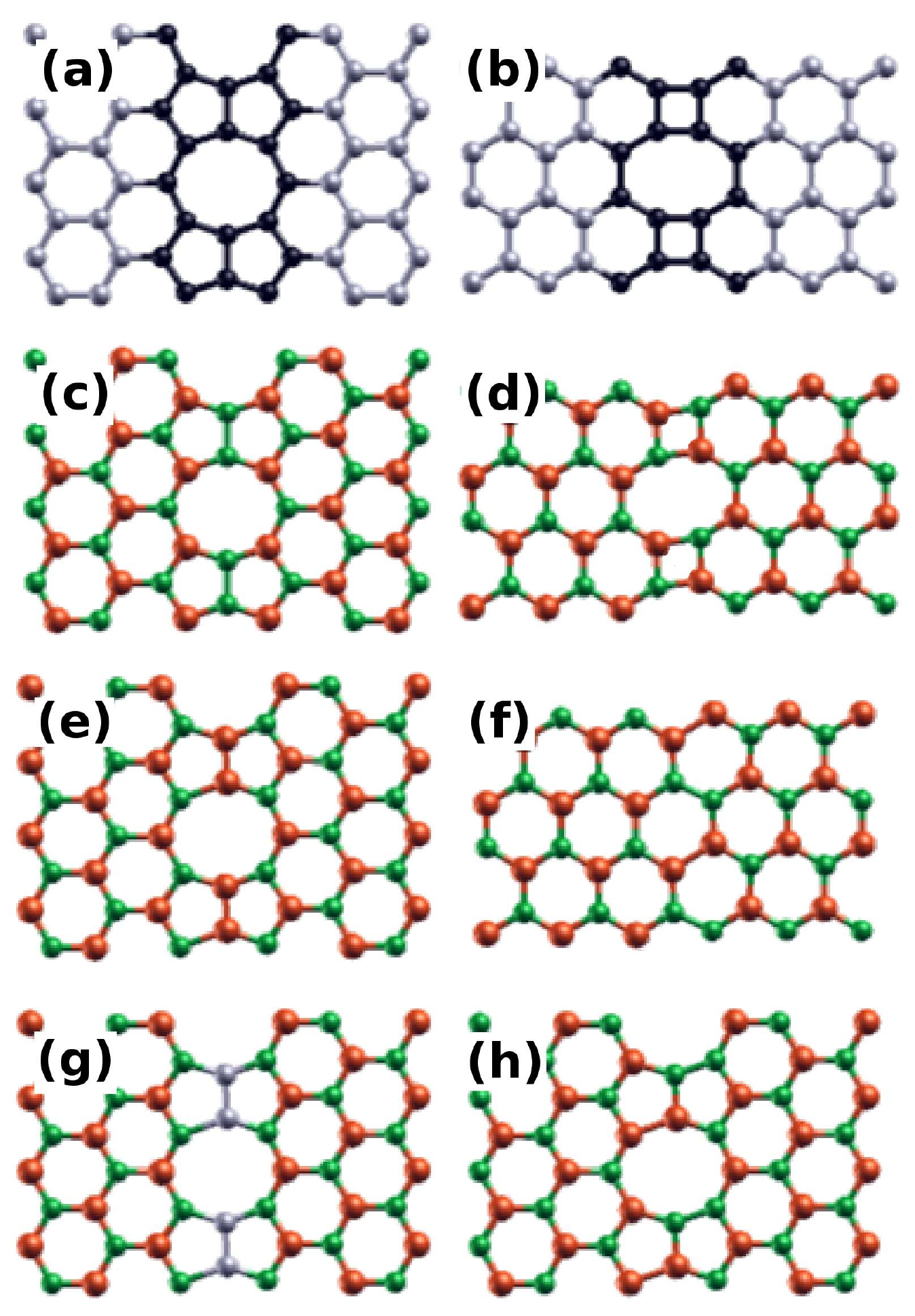}
\caption{Structures of grain boundaries (GB) and antiphase boundaries
(APB) in monolayer graphene and boron nitride. Boron, nitrogen, and
carbon atoms are shown by orange, green, and grey circles,
respectively. (a) Z-GB: a graphene GB with zigzag chirality. (b) A-GB:
a graphene GB with armchair chirality. (c) NZ-APB: a zigzag APB in BN
with N-rich core. (d) A-APB: an armchair APB in BN with stoichiometric
core. (e) BZ-APB: a zigzag APB in BN with B-rich core. (f) HA-APB: an
armchair APB in BN with stoichiometric core consisting of hexagons
with homopolar bonds. (g) CZ-APB: a zigzag APB in BN with carbon-doped
core. (h) Z-GB: a zigzag GB in BN with stoichiometric core. Core atoms
for zigzag and armchair defects are shown by darker circles in (a) and
(b), respectively.}
\label{fig1}
\end{figure}
Many proposed applications of nanomaterials require the ability to
control their electronic properties. In particular, graphene and boron
nitride (BN) in the two-dimensional (2D) monolayer form have become an
important subject of research, owing to their mechanical strength and
a rich variety of physical phenomena connected to their electronic
structure.~\cite{rmp} The introduction of structural defects presents
an alternative for manipulating the electronic and magnetic properties
in these materials.~\cite{rmp,banhart,umesh,lahiri,si-nl,xli,pochet}
In graphene grown on Ni(111) substrates, a translational grain
boundary (GB) has been theoretically proposed~\cite{umesh} and
recently observed in STM experiments,~\cite{lahiri} and the occurrence
of magnetism for the quasi-one-dimensional electronic states
introduced by this defect has been suggested by {\it ab initio}
calculations.~\cite{si-nl} This GB in graphene lies along the zigzag
direction and arises due to the presence of two possible stackings of
the graphene monolayer with respect to the Ni(111) substrate, which
leads to the possibility of domains related by a relative translation,
with the GB emerging as the boundary between two such
domains.~\cite{lahiri} In the case of monolayer BN grown on Ni(111),
the same stacking mechanism holds,~\cite{auwarter} and the possibility
of engineering smaller band gaps in this large-band-gap material by
the introduction of this zigzag-direction boundary has been recently
considered.~\cite{xli}

In this work, we introduce a low-energy stoichiometric model for an
armchair-direction antiphase boundary (APB) in monolayer
BN,~\cite{lidia-aps} based on a structural pattern recently observed
experimentally in irradiated graphene.~\cite{kotakoski} Furthermore,
we investigate the electronic properties and compare the stability of
several different models for extended one-dimensional (1D) defects in
monolayer BN and graphene, including the aforementioned zigzag and
armchair boundaries.

Our {\it ab initio} calculations indicate that in graphene the zigzag
GB is more stable than the armchair GB, while in BN the relative
stability of APBs with zigzag and armchair chiralities depends on the
chemical potentials of the B and N species in the synthesis process,
with nitrogen-rich zigzag APBs becoming more stable than
stoichiometric armchair APBs only when defects are formed under
nitrogen-rich conditions. Moreover, in the presence of carbon dopants,
a C-doped zigzag APB becomes the most stable boundary for the last
one-third of the interval of realistic B and N chemical potentials
values, comprising the last part of the intrinsic region plus the
N-rich region. Furthermore, we find that in graphene the armchair GB
introduces weaker resonances near the Fermi level (FL), associated to
defect states that are only partially confined to the defect core, in
contrast with the strongly localized states characteristic of the
zigzag GB.~\cite{si-nl} In the case of BN, we find that the armchair
APB introduces flat bands near the bulk band edges which are weakly
confined to the defect core, while the B-rich, the N-rich, and the
C-doped zigzag APBs lead to the formation of electronic bands that are
deep in the bulk band gap and strongly localized on the atoms at the
defect core.

In graphene, the translational GB observed in Ref.~\onlinecite{lahiri}
is obtained by cutting a graphene sheet along the zigzag direction,
displacing the two halves by one-third of the lattice period in the
direction perpendicular to the cut, and inserting carbon dimers with
their common bond oriented along the cut, generating a line of
pentagon-pentagon-octagon units, as shown in Fig.~\ref{fig1}(a). Being
oriented along the zigzag direction, this boundary is labeled Z-GB in
our discussion.

In the present study, we examine the armchair-direction counterpart of
the Z-GB in graphene. The armchair grain boundary, which we label
A-GB, can be obtained by cutting a graphene sheet in the armchair
direction and translating one side of the sheet with respect to the
other side by half the lattice period along the armchair direction,
generating a line defect that contains squares and octagons
alternately arranged in its core, as shown in
Fig.~\ref{fig1}(b). Small finite segments of this core structure
have been observed in graphene as result of reconstruction after
electronic-beam irradiation.~\cite{kotakoski} Chiral GB geometries may
also be built by combining these two basic structures.

In a binary system like monolayer BN, inversion symmetry is absent and
an APB is formed at the interface of two domains with opposite
assignments of B and N atoms to the two triangular sublattices of the
BN honeycomb lattice. In this material, the geometries of the graphene
Z-GB and A-GB translate into APBs: the core of the armchair boundary,
which we label A-APB, is naturally stoichiometric (having the same
number of N and B atoms) as shown in Fig.~\ref{fig1}(d), while for
the zigzag APB the core is either N-rich or B-rich, if one adopts the
criterion of minimizing the number of homopolar (N-N or B-B) bonds.
In our discussion, the N-rich version of the zigzag boundary, shown in
Fig.~\ref{fig1}(c), is labeled NZ-APB; the boron-rich one, shown in
Fig.~\ref{fig1}(e), is labeled BZ-APB; a carbon-doped version
obtained by replacing the B$_2$ dimers at the center of the BZ-APB
core with C$_2$ dimers, shown in Fig.~\ref{fig1}(g), is labeled
CZ-APB. A model for a stoichiometric Z-GB in BN (not an APB in this
case) is also possible, at the cost of introducing an extra homopolar
bond per defect unit, as shown in Fig.~\ref{fig1}(h). We also
consider a stoichiometric APB interface in BN containing homopolar
bonds without topological defects, shown in Fig.~\ref{fig1}(f).

In our study, we employ a first principles approach based on Kohn-Sham
density functional theory (KS-DFT),~\cite{ks} as implemented in the
SIESTA code.~\cite{siesta} All calculations are performed using the
generalized-gradient approximation (GGA) for the exchange-correlation
term.~\cite{gga} Interactions between valence electrons and ionic
cores are described by Troullier-Martins pseudopotentials.~\cite{tm} A
double-$\zeta$ pseudoatomic basis set augmented with polarization
orbitals is employed, with an energy cutoff of 0.01 Ry. Structural
optimization is performed until the residual force on each atom is
less than 0.04 eV/\AA. Supercells in our study are periodic in the
monolayer plane and large vacuum regions are included to impose
periodic boundary conditions in the perpendicular direction. Ribbon
geometries are described below. Convergence tests were performed and
supercell sizes were chosen such that interactions between defects in
the plane and between each layer and its periodic images were
negligible.

We seek to compare the relative stability of the above 1D defects in
graphene and BN. In the case of graphene, the formation energy of the
GBs per unit length $E_{f}^{GB}$ is given by
\begin{equation}
 E_{f}^{GB} = \frac{E_{tot}^{GB}(N)-N\mu_{graph}^{bulk}}{\ell}
\label{ef-gb}
\end{equation}
where $E_{tot}^{def}(N)$ is the total energy of the $N$-atom supercell
containing a GB, $\mu_{graph}^{bulk} =$-154.532~eV is the bulk
chemical potential of graphene, obtained as the total energy per atom
in a pristine graphene calculation, and ${\ell}$ is the length of the
supercell along the defect direction.

Our calculated values for $E_f^{Z-GB}$ and $E_f^{A-GB}$ in graphene
are included in Table~I. In this material, the Z-GB is more stable
than the A-GB by 0.25 eV/\AA\ due to the smaller bond-length and
bond-angle distortions from the ideal bulk values ($d_{bulk} =
1.442\;\!$\AA\ and $\theta_{bulk} = 120^\circ$ in our calculations)
incurred in the pentagon-pentagon-octagon core of the Z-GB, when
compared with the tetragon-octagon core of the A-GB. Average bond
lengths and bond angles, as well as maximum and minimum values and
standard deviations for these quantities are included in Table~I for
the Z-GB and the A-GB. While average values are similar for both bond
lengths and bond angles, deviations from the bulk reference values are
larger in the A-GB core.

This indicates that the nature of the energy difference between the
A-GB and the Z-GB in graphene is essentially elastic. Indeed, a
Keating-model calculation with a Keating potential fitted for diamond
carbon~\cite{martin} predicts the elastic energy of the A-GB to be
about twice that of the Z-GB, in qualitative agreement with our {\it
ab initio} results. A more quantitative agreement would require
fitting the Keating potential to the graphene bonding environment.
\begin{table}[h]
\centering
\begin{tabular}{|l|c|c|c|c|c|c|c|c|c|}
\hline 
&$E_{f}$ &$\bar{d}$ &$d_{max}$
&$d_{min}$ &$\sigma_d$ &$\bar{\theta}$ &$\theta_{max}$
&$\theta_{min}$ &$\sigma_{\theta}$
\\   \hline 
Z-GB       &    0.48              & 1.45 & 1.48 &1.41 & 0.02 & 120$^\circ$ 
& 141.7$^\circ$ & 104.6$^\circ$
&11.1$^\circ$   \\ \hline 
A-GB       &    0.73              & 1.44 & 1.52 &1.40 & 0.04 & 120$^\circ$ 
& 147.9$^\circ$ & 90$^\circ$
&16.8$^\circ$   \\ 
\hline 
\end{tabular}
\caption{Formation energies ($E_f$ in eV/\AA), and average, maximum,
minimum, and dispersion values for bond lengths ($d$ in \AA) and bond
angles ($\theta$) for zigzag and armchair grain boundaries in
graphene.}
\label{table-GB}
\end{table}

\begin{figure}
\centering \includegraphics[width=8cm]{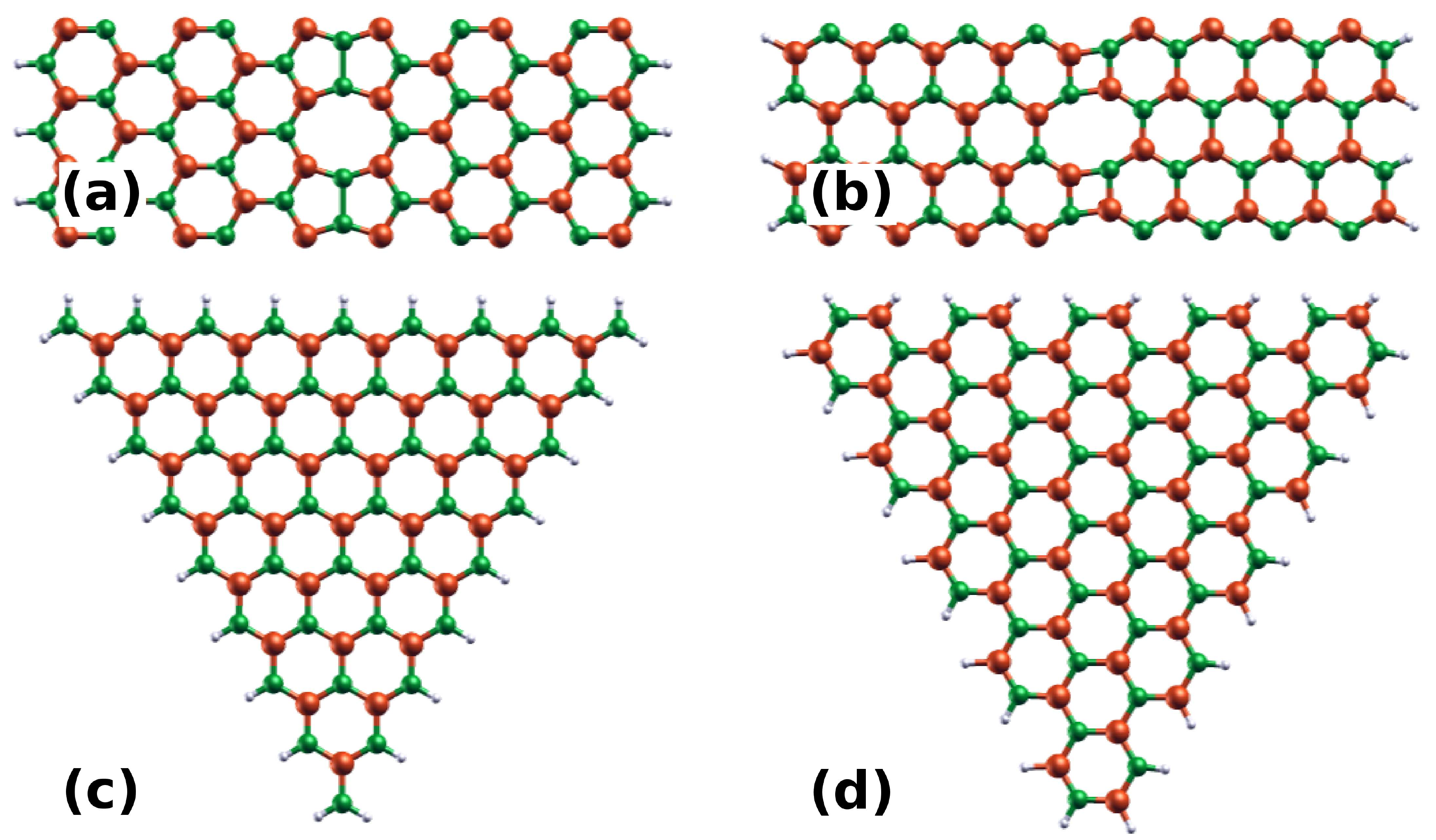}
\caption{Ribbon and triangle geometries for computation of
extended-defect energies in monolayer boron nitride. (a) Ribbon with N-rich
zigzag antiphase boundary in the middle and N-rich zigzag edges. (b)
Ribbon with armchair boundary in the middle and armchair edges. (c)
Triangle with the same N-rich zigzag edges as ribbon in (a). (d)
Triangle with the same armchair edges as ribbon in (b).}
\label{fig2}
\end{figure}

In BN, the lack of inversion symmetry means that in a periodic
supercell calculation for non-stoichiometric boundaries, such as the
NZ-APB and its BZ-APB partner, both boundaries are included in the
periodic cell, hence only the sum of the formation energies of the two
1D defects can be extracted from such calculation. Obtaining formation
energies of individual boundaries requires using BN ribbons,
containing a single boundary in the middle and hydrogen-saturated
edges. The ribbons are finite in the direction perpendicular to the 1D
boundary and periodic in the parallel direction. Figure~\ref{fig2}(a)
shows the BN ribbons employed for the NZ-APB calculation, and
Fig.~\ref{fig2}(b) shows the ribbon employed in the A-APB
calculations. The ribbon employed for BZ-APB calculations (not shown
in Fig.~\ref{fig2}) is obtained from the NZ-APB one by swapping the
B and N atoms.

Stoichiometry is determinant for the stability of the various APBs and
GBs in BN, since the chemical potentials of the B and N species will
depend on the growth conditions, i.e., on the sources of B and N atoms
in the synthesis process. The formation energy of the BN ribbons
depends on the ribbon stoichiometry and is thus a function of the
chemical potentials for B, N, and H. Assuming that defect formation
occurs in equilibrium with a bulk BN monolayer, we impose that $\mu_B
+ \mu_N = \mu_{BN}$, where $\mu_{BN}$ is the total energy per BN pair
for a pristine BN monolayer, and explore the APB formation energy
$E_f^{APB}$ as a function of $\mu_B$ and $\mu_N$ by defining two
limiting chemical potential environments: in the B-rich case $\mu_B$
is obtained from the total energy per atom for the $\alpha$-boron bulk
phase ($\alpha-B$), and in the N-rich situation $\mu_N$ is
obtained from the total energy per atom of an $N_2$ molecular crystal
calculation. We write
\begin{eqnarray}
&\;\!&\mu_N + \mu_B = \mu_{BN} = -350.187~{\rm eV}\; ;\nonumber \\
&\;\!&\mu_N^{max} = \frac{E_{tot}^{N_2}}{2} = -270.290~{\rm eV } \;\;\;\left({\rm
N-rich}\right)\;;\\ 
&\;\!&\mu_N^{min} = \mu_{BN} -
\frac{E_{tot}^{\alpha\!B}}{12} = -273.712~{\rm eV } \;\;\;\left({\rm
B-rich}\right)\;.\nonumber
\label{eq-mu}
\end{eqnarray}

We emphasize that our range of values for $\mu_B$ and $\mu_N$ are
determined for a condition of equilibrium with solid-state phases of
BN and N (or B). These physical constraints, previously considered in
calculations of defects in bulk BN,~\cite{walter} BCN
monolayers,~\cite{sergio1} BN fullerenes,~\cite{si1,si2} and BN
nanocones,~\cite{sergio2} have not been considered in a recent
calculation.~\cite{triang} As shown in the following, the
determination of physically acceptable ranges for $\mu_N$ and $\mu_B$
is crucial to determine which extended defect structure is the most
stable under N-rich, intrinsic, or B-rich conditions.

Given the chemical potentials in Eq.~2, we define the formation
energy of the BN ribbons $E_{f}^{rib}$, which includes the formation
energies of the APB and of the edges, as follows:
\begin{eqnarray}
E_{f}^{rib} &=& \frac{E_{tot}^{rib}-N_B \mu_B -
N_N \mu_N-N_H \mu_H}{\ell} \label{ef-rb1}\\ 
E_{f}^{rib} &=& E_f^{APB}+ 2  E_f^{edge}.
\label{ef-rb2}
\end{eqnarray}
where $E_{tot}^{rib}$ is the calculated total energy of the ribbon,
$\ell$ is the length of the ribbon along the APB direction, $N_B$,
$N_N$, and $N_H$ are the numbers of boron, nitrogen, and hydrogen
atoms in the ribbon, and $\mu_B$, $\mu_N$, and $\mu_H$ are the
respective chemical potentials.

In order to obtain $E_f^{APB}$ from Eq.~\ref{ef-rb2} above, we follow
the procedure from Ref.~\onlinecite{triang} and consider BN triangles in
which the three hydrogen-saturated edges are the same as those in the
corresponding ribbons, as shown in Fig.~\ref{fig2}. The formation
energy of an $N$-atom triangle $E_{f}^{\triangle}$ is defined
similarly to Eq.~\ref{ef-rb1}, and can be decomposed into three
components: a bulk one that scales with the area of the triangle
($\propto\!N$), an edge one that scales with the edge length
($\propto\!N^{1/2}$), and a vertex component that does not scale with
the size of the triangle. Since the bulk of the triangles is composed
of BN units, and BN bulk is our reference chemical potential
(c.f. Eqs.~2 and \ref{ef-rb1}), the bulk component of
$E_{f}^{\triangle}$ vanishes by definition. It is then possible to
obtain the edge energy per edge unit, by considering triangles of
different sizes, and fitting $E_{f}^{\triangle}$ to a linear form:
\begin{equation}
E_{f}^{\triangle} =
\lambda_f^{edge} n_{edge}+E_f^{vtx}\;;
\end{equation}
where $\lambda_f^{edge}$ is the edge energy per edge unit, $n_{edge}$
is the number of edge units in the triangle, which for zigzag-edge
triangles and ribbons is the number of N (B) atoms saturated with one
hydrogen in Fig.~\ref{fig2}, and for armchair-edge triangles and
ribbons is the number of boat-like BN units at the edges. $E_f^{vtx}$
is the contribution from the three vertices of the triangles.
\begin{figure}[]
\centering \includegraphics[width=6cm]{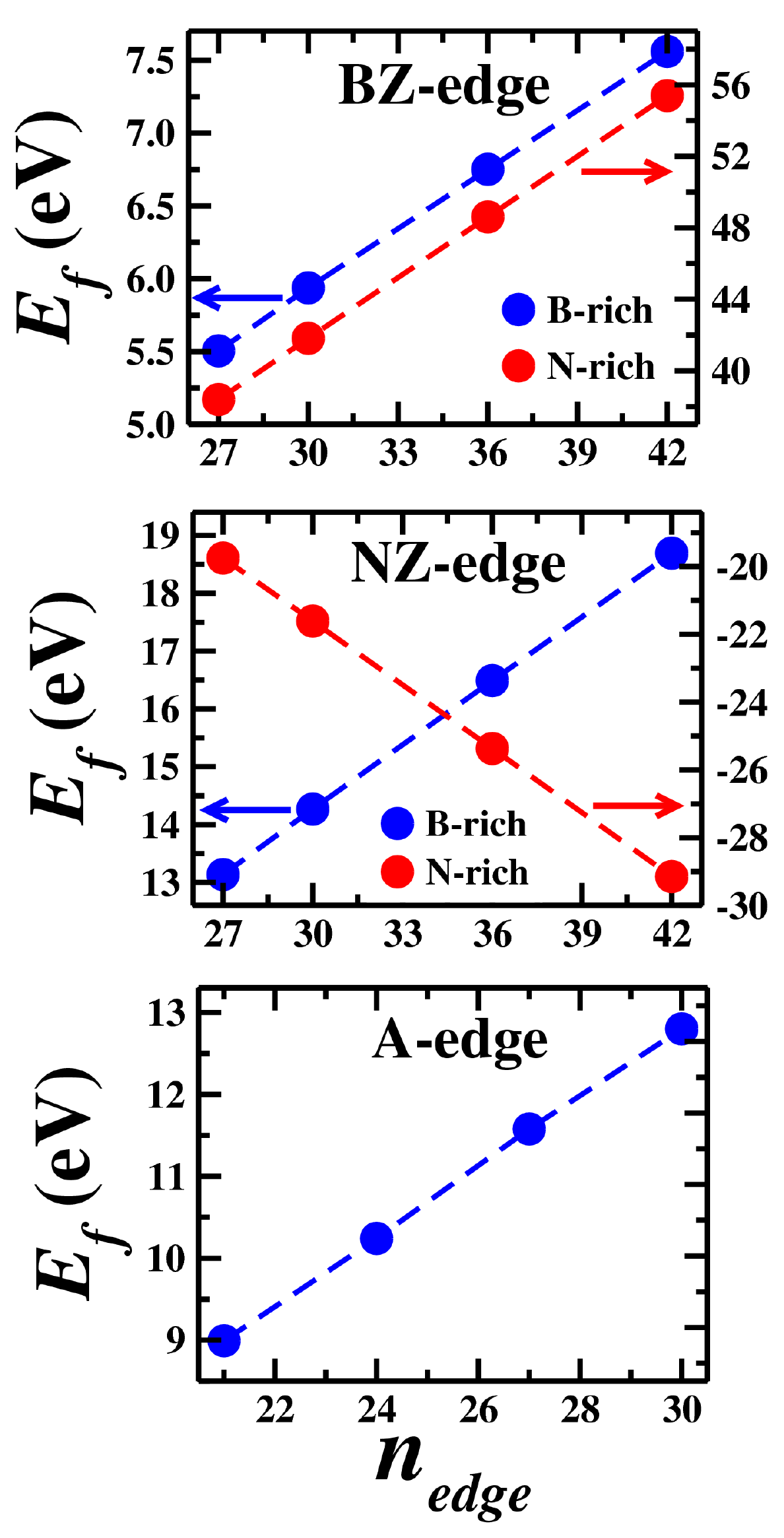}
\caption{Formation energy of BN triangles as a function of the number
of edge units. The upper (middle) panel shows the energies of the
NZ-edge (BZ-edge) for N-rich and B-rich environments. The lower panel
shows the energy of the stoichiometric A-edge.}
\label{fig3}
\end{figure}

The fittings we obtain for the energies of the triangles shown in
Fig.~\ref{fig2}(c), with N-rich zigzag edges (NZ-edge), and for the
triangles with B-rich zigzag edges (BZ-edge), under B-rich and N-rich
environments, are shown in the upper and middle panels of
Fig.~\ref{fig3}, respectively. From the slope of the curves in
Fig.~\ref{fig3} we obtain the NZ-edge and BZ-edge energies included in
Table~\ref{triang}. In order to check the consistency of our approach,
we also compute the edge energy for triangles shown in
Fig.~\ref{fig2}(d) with armchair edges (A-edge), and $E_f^{A-APB}$
from Eq.~\ref{ef-rb2} for the ribbon containing an A-APB and armchair
edges, shown in Fig.~\ref{fig2}(b). The fitting of the energies of the
armchair-edge triangles is shown in the lower panel in
Fig.~\ref{fig3}, and the energy of the A-edge is also included in
Table~\ref{triang}. The A-edge is stoichiometric, hence its energy is
independent of the B and N chemical potentials.
\begin{table}[h]
\centering
\begin{tabular}{|c|c|c|c|c|}
\hline 
         &\multicolumn{2}{|c|}{$\lambda_f^{edge}$ (eV)}
         &\multicolumn{2}{|c|}{$E_f^{vtx}$ (eV)} \\ \hline
A-edge   &\multicolumn{2}{|c|}{0.43} & \multicolumn{2}{|c|}{0.06} \\ \hline
         &B-rich & N-rich &B-rich &N-rich \\ \hline 
NZ-edge  &0.37  &-0.59   &3.15   &-2.56   \\ \hline
BZ-edge  &0.14   & 1.09   & 1.83  & 7.54  \\ \hline 
\end{tabular}
\caption{Edge energy per edge unit $\lambda_{form}^{edge}$ and vertex
energy $E_f^{vtx}$, obtained from fitting of triangle
energies to Eq.~5, for the armchair edge (A-edge), the N-rich zigzag edge
(NZ-edge), and the B-rich zigzag edge (BZ-edge).}
\label{triang}
\end{table}

It is worth commenting on the negative slope of $E_f$ for the NZ-edge
under an N-rich environment. It indicates that the reaction by which
hydrogen saturates the edge is exothermic and is consistent with
experimental observations of a very high stability for
nitrogen-terminated edges in BN islands.~\cite{kim,jin,meyer} Indeed,
the strong tendency of BN patches to display a triangular shape is
connected with the stability of the NZ-edges.

\begin{figure}
\centering \includegraphics[width=8cm]{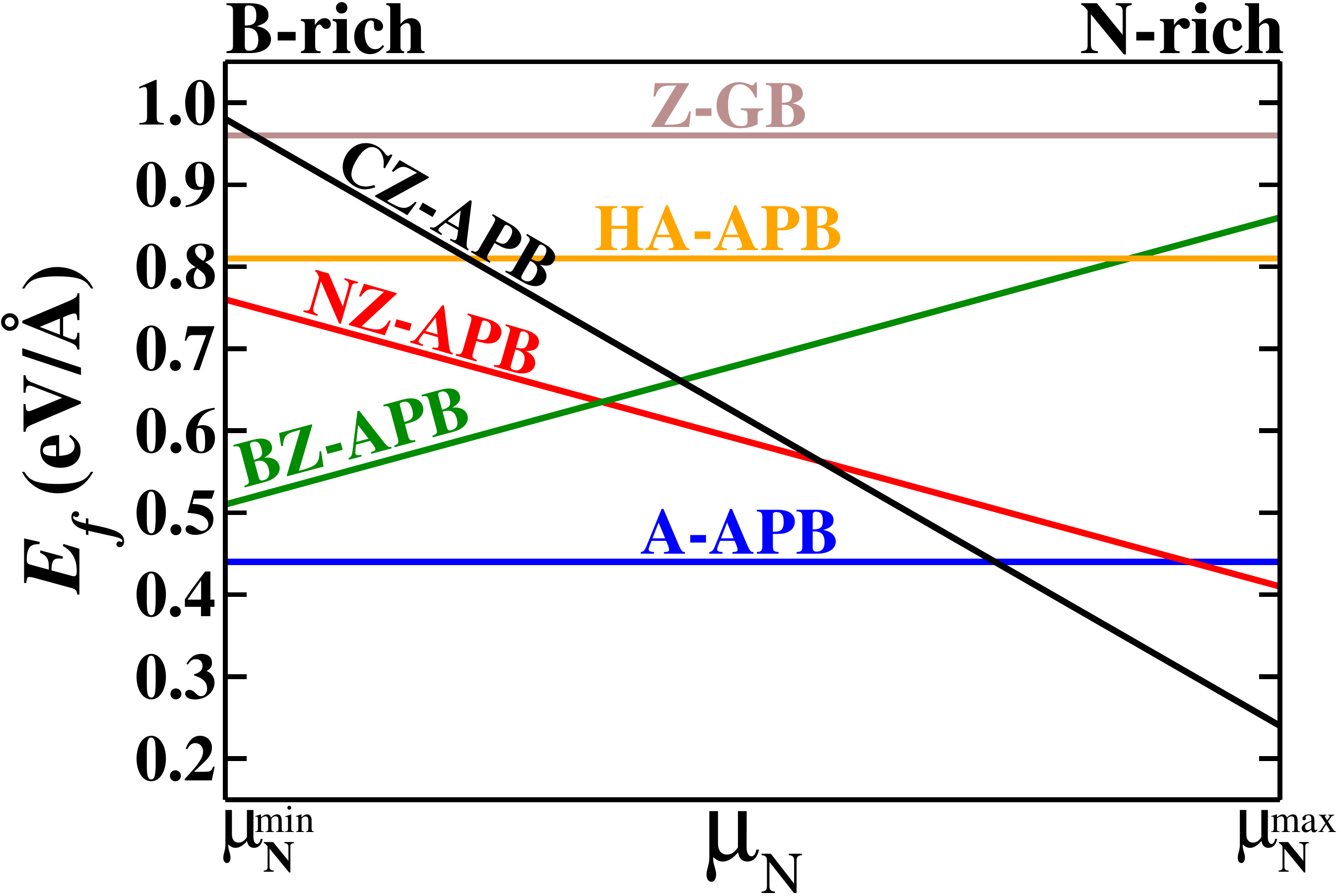}
\caption{Formation energy $E_f$ of grain boundaries and antiphase
boundaries in boron nitride as a function of the nitrogen chemical
potential $\mu_N$. The maximum and minimum values of $\mu_N$ are given
in the text (see Eq.~2)}
\label{fig4}
\end{figure}

Having obtained $E_f^{edge}$ from the above procedure, we can obtain
$E_f^{APB}$ from Eq.~\ref{ef-rb2}. The results for the five APB
models, as well as $E_f^{Z-GB}$ for the Z-GB in BN, as functions of
$\mu_N$, are shown in Fig.~\ref{fig4}. For the stoichiometric models
$E_f^{APB}$ is independent of $\mu_N$. The consistency of the above
procedure is attested by the fact that $E_f^{APB}$ values obtained
using supercells and the H-terminated ribbons agree to within 0.6\% in
all cases, as included in Table~\ref{e-bn} (for the NZ-APB NZ-APB we
compare the sum of the formation energies).

It is evident that the A-APB is the more stable undoped boundary
across $\sim$80\% of the range of chemical potentials we consider,
with the NZ-APB becoming the most stable one in the N-rich limit of
$\mu_N$. In the presence of a carbon source, the carbon-doped CZ-APB
is the least stable in a B-rich environment, but becomes the most
stable boundary for about one third of the interval of $\mu_N$ values,
from the end of the intrinsic regime through the N-rich limit. We note
that the elastic-energy cost of the A-APB should be even higher than
that for the A-GB in graphene, because in BN the bond angles at the
core of the defect depart even more strongly from the ideal bulk value
of 120$^\circ$ than in graphene, as can be seen in
Fig~\ref{fig1}. Unlike the case in graphene, however, BN strongly
prefers even-membered topological defects in order to avoid the
energetically expensive homopolar bonds, except in the N-rich limit
where N-N bonds become more favorable.~\cite{si1}
\begin{table}[h]
\centering
\begin{tabular}{|c|c|c|}
\hline
         &\multicolumn{2}{|c|}{$E_f$ (eV/\AA{})}         \\  \hline
Boundary & supercell           & ribbon                  \\  \hline
A-APB    & 0.44                & 0.44     \\  \hline
HA-APB   & 0.81                & 0.81     \\  \hline
Z-GB     & 0.96                & 0.95     \\  \hline
NZ-APB + BZ-APB & 1.27           & 1.27     \\  \hline
                & N-rich - ribbon   & B-rich - ribbon  \\  \hline
NZ-APB          & 0.41   & 0.76                   \\  \hline
BZ-APB          & 0.86   & 0.51                   \\  \hline
CZ-APB          & 0.19   & 0.98                   \\  \hline
\end{tabular}
\caption{Comparison of formation energy values $E_f$ of grain
boundaries and antiphase boundaries in monolayer BN computed with
supercells and ribbon geometries. Supercells are stoichiometric and
corresponding $E_f$ values do not depend on chemical potentials of B
and N. For the ribbons, values for N-rich and B-rich environments are
included.}
\label{e-bn}
\end{table}

We can use a combination of two models to understand qualitatively the
ordering of $E_f$ values in Fig.~\ref{fig4}. The idea is to divide
the contributions to the formation energies of the boundaries in two
components, an elastic one $E_{el}$, which we estimate by employing a
Keating potential fitted for cubic BN,~\cite{bn-keat} and a chemical
energy $E_Q$ that is computed using the bond-energy model developed in
Ref.~\onlinecite{sergio1} to account for the energetics of CBN sheets of various
stoichiometries. In this latter model, two-atom bonds are assigned
bond-energy values, which reflect the average energy of each type of
bond across the various BCN sheets included in the fitting procedure.

The bond-energy values derived in Ref.~\onlinecite{sergio1} are
$\varepsilon_{CC} = -103.24$~eV, $\varepsilon_{BN}\!=\!-116.73$~eV,
$\varepsilon_{CN}\!=\!-141.67$~eV, $\varepsilon_{BB}\!=\!-50.40$~eV,
and $\varepsilon_{NN} = -178.49$~eV, for carbon-carbon,
boron-nitrogen, carbon-nitrogen, boron-boron, and nitrogen-nitrogen
bonds, respectively. Within the model, the values of $E_Q$ for one
period of either the H-APB or the Z-GB, as well as the sum of $E_Q$
values for one period of the NZ-APB and the BZ-APB, should all be
$\sim$4.62 eV higher than the value for the A-APB, i.e., $E_Q^{H-APB}
= E_Q^{Z-GB} = E_Q^{(NZ+BZ)-APB} = E_Q^{A-APB} + 4.62~$eV. Using the
BN Keating potential from Ref.~\onlinecite{bn-keat}, we compute the values of
$E_{el}$ for the five different boundaries. Results for $E_{el}$,
$E_{Q}$, and $E_f = E_Q + E_{el}$ for the A-APB, H-APB, Z-GB, and the
sum of the values for the NZ-APB and BZ-APB, computed using this
scheme, are included in Table~\ref{e-model}.
\begin{table}[h]
\centering
\begin{tabular}{|c|c|c|c|}
\hline
Boundary            & $E_Q$(eV/\AA) & $E_{el}$ (eV/\AA) & $E_f$(eV/\AA)   \\  \hline
A-APB               & 0.00           & 1.02             & 1.02            \\  \hline
NZ-APB + BZ-APB     & 0.92           & 1.08             & 2.00            \\  \hline
HA-APB              & 1.06           & 0.18             & 1.24            \\  \hline
Z-GB                & 0.93           & 0.70             & 1.63            \\  \hline
\end{tabular}
\caption{Formation energy of grain boundaries and antiphase
boundaries in monolayer BN from combination of elastic energy and
chemical-bond energy models.}
\label{e-model}
\end{table}

Despite the fact that this simplified partition of $E_f$ fails in
describing, even qualitatively, the energy difference between the
A-APB and the average of the NZ-APB and BZ-APB, it correctly predicts
that the Z-GB has the highest value of $E_f$, followed by the
H-APB. This modeling is useful because it allows us to establish that
$E_{Q}$ is the predominant factor in determining the relative stability
of the various boundary models in our study. Note the very low value
of $E_{el}$ for the H-APB, with a defect core consisting of hexagons
only, which is offset by the high energy cost of two ``wrong'' bonds
per defect unit in the core. On the other hand, the highest value of
$E_{el}$ is found for the A-APB, being associated with the strong
departure from the bulk bond lengths and bond angles in the fourfold
and eightfold rings at the core of this boundary. However, this is
compensated by the fact that the chemical energy cost for the A-APB is
very low ($E_Q^{A-APB} = 0$ within the simplified bond-energy model
above). 

Again, a quantitative agreement between the above modeling and the
{\it ab initio} results would require fitting both the Keating
potential and the bond-energy model to bonding environments that are
more similar to those in the boundary geometries in our study.  This
analysis also provides a qualitative explanation for one of the
ingredients that determines the stability of the carbon-doped CZ-APB
boundary: because this structure is obtained from the BZ-APB by
replacing the B$_2$ dimer in the center of the defect core with a
C$_2$ dimer, the elastic energy is reduced because the length of the
C-C bonds is similar to the length of the BN bulk matrix, unlike the
longer B$_2$-dimer bond. The other ingredient is electronic, and is
determined by stoichiometry (i.e., the chemical potentials) and the
fact that in low coordination bonding environments a B-B bond is
strongly less favorable than a C-C bond.

\begin{figure}
\centering \includegraphics[width=8cm]{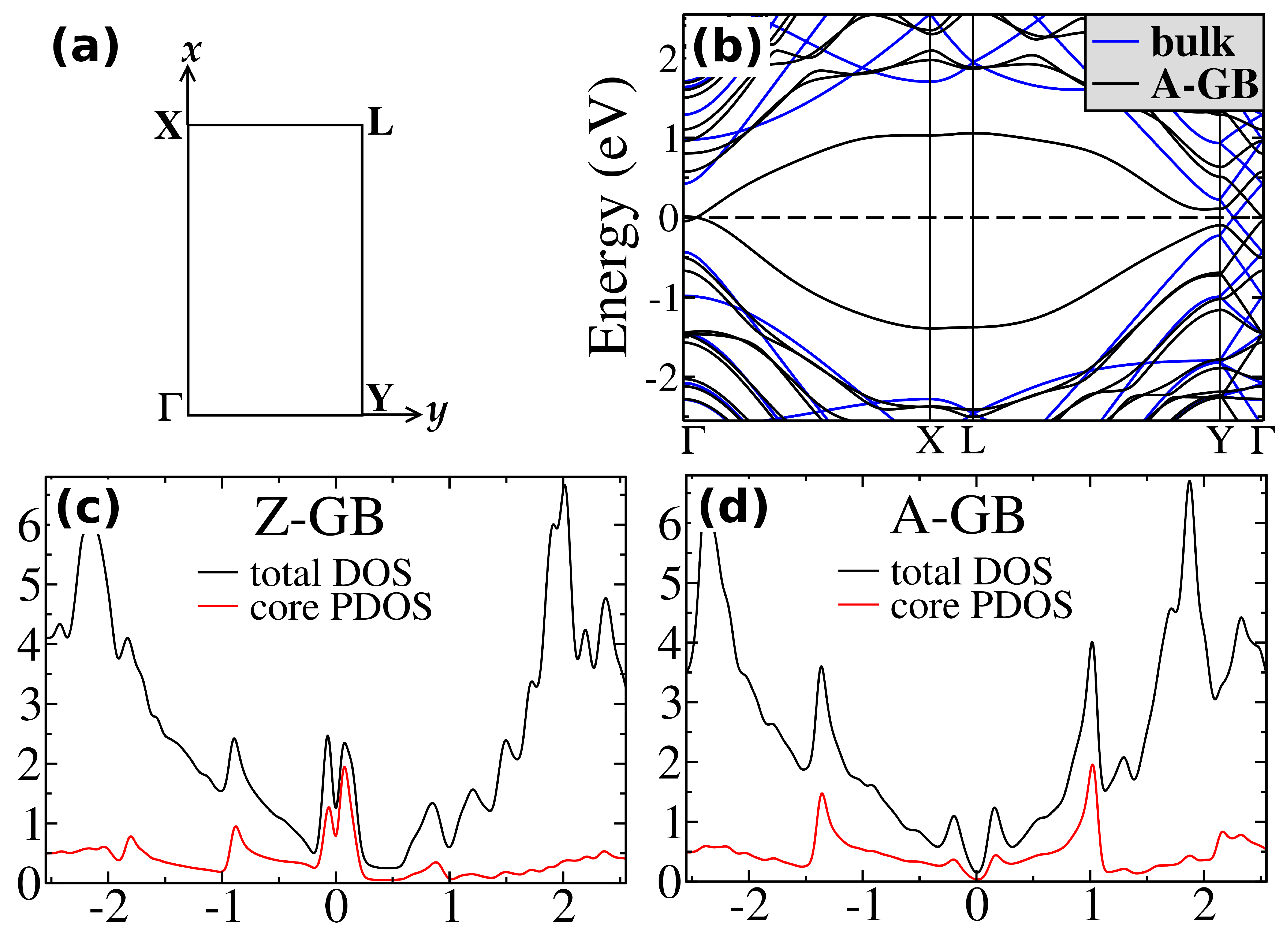}
\caption{Band structure and density of states for the A-GB and Z-GB in
graphene. (a) Brillouin zone corresponding to the supercell
calculations in the present study. (b) Black curves show the A-GB
supercell band structure. Blue curves show bulk bands folded onto
defect supercell.  (c) and (d) show total density of states (DOS) and
the DOS projected on the core atoms for the Z-GB and A-GB, respectively.}
\label{fig5}
\end{figure}

Let us now examine the electronic structure of the graphene
boundaries. The electronic structure of the Z-GB in graphene has been
discussed in Ref.~\onlinecite{si-nl}, where the appearance of a
highly-localized quasi-1D state, introduced by this boundary in the
density of states (DOS) of graphene, has been shown to lead to a
magnetic instability. [The Z-GB is labeled GB$(2,0)|(2,0)$ in the
notation employed in Ref.~\onlinecite{si-nl}.] Figures~\ref{fig5}(b)
and (d) show the electronic band structure and the DOS for the A-GB in
graphene, and the DOS for the Z-GB is shown in Fig.~\ref{fig5}(c) for
comparison. The Brillouin zone corresponding to all defect supercells
in this work is shown in Fig.~\ref{fig5} (a). The $\Gamma$-X and Y-L
lines are parallel to the defect direction (the $x$-axis of the
supercell) in all cases. In order to identify the boundary-related
electronic bands, and also to examine the degree of localization of
the corresponding states on the core of the boundaries, we project the
DOS onto the orbitals centered on the core atoms in each case. The
core atoms for each defect geometry are shown by darker circles in
Fig.~\ref{fig1}(a) and (b), and we use the same definition of defect
core for the graphene and BN boundaries.

From the A-GB band structure, we can see that this boundary introduces
only small electron and hole pockets near the Fermi level (FL), that
show as two weak resonances within $\pm$0.2 eV from the FL in the
DOS. The degree of localization of these electronic states on the
eight atoms located at the A-GB core is much weaker than what is found
for the magnetic resonance in the Z-GB case.~\cite{si-nl} In
the latter case, nearly 90\% of the DOS derives from the ten core
atoms. This is shown by the partial DOS (PDOS) for the core atoms as
red curves in Fig.~\ref{fig5}(b) and (d). In the A-GB case, these
resonances are much more spread out into the bulk of the cell, and the
contribution from the core atoms is much smaller than in the Z-GB. The
A-GB also gives rise to stronger resonances at $\pm$1.0 eV from the
FL, connected to the flat portion of the defect bands seen in the band
structure in Fig.~\ref{fig5}(a)].

\begin{figure}
\centering \includegraphics[width=8cm]{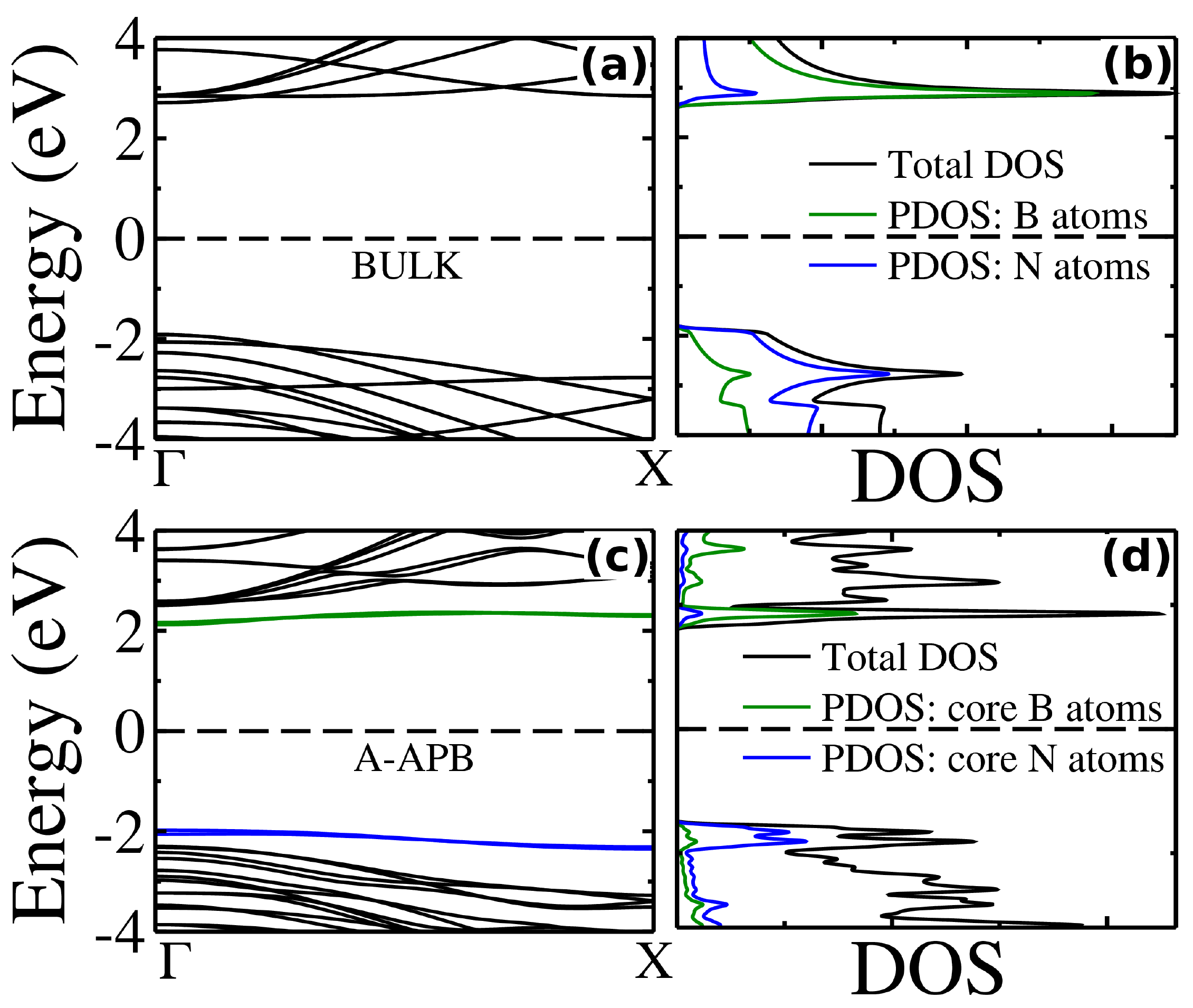}
\caption{(a) and (b) Band structure and density of states for bulk BN.
Blue and green curves in (b) show the contributions of orbitals of the
N and B atoms to the total DOS, respectively. (c) and (d) Band
structure and density of states for the A-APB in BN. The contribution
of the core-atom orbitals to the total DOS is shown by blue (N
orbitals) and green (B orbitals) curves in (d). Defect bands in the
gap are shown by green and blue curves, according to the dominant
atomic-orbital contribution in each case.}
\label{fig6}
\end{figure}

In the case of BN, we concentrate on the low-energy A-APB, NZ-APB,
BZ-APB, and CZ-APB models. In Fig.~\ref{fig6}(a) and (b) we show the
band structure and the density of states (DOS) for a pristine bulk BN
monolayer. For ease of comparison, the bulk calculation was performed
for a supercell of nearly the same dimensions as those employed for
the BN boundaries. For our discussion, the important features of the
bulk electronic structure are the size of the gap ($\sim$4.7 eV within
the GGA-DFT scheme) and the composition of the electronic states at
the band edges: the top of the valence band is mainly composed of
nitrogen $p_z$ orbitals while the bottom of the conduction band
derives from the boron $p_z$ orbitals.

The electronic bands and the corresponding DOS for the A-APB are shown
in Fig.~\ref{fig6}(c) and (d). The A-APB introduces one set of two
bands near each of the band edges, that show little dispersion and
retain the character of the corresponding parent bulk bands. The
acceptor bands near the valence-band maximum (VBM) are composed
primarily of nitrogen $p_z$ orbitals, while the donor bands near the
conduction-band minimum (CBM) consist of boron $p_z$ orbitals, with
$\sim$40\% of the DOS concentrated on the core atoms in both cases, as
shown by the core-projected PDOS curve in Fig.~\ref{fig6}. Both sets
of bands are shallow ($\sim$0.2 eV split from the corresponding band
edges), and the lack of sizeable dispersion indicates very large
effective masses and low mobilities of carriers, should doping of A-GB
defect bands become feasible.

\begin{figure}
\vspace{0.1cm}
\centering \includegraphics[width=8cm]{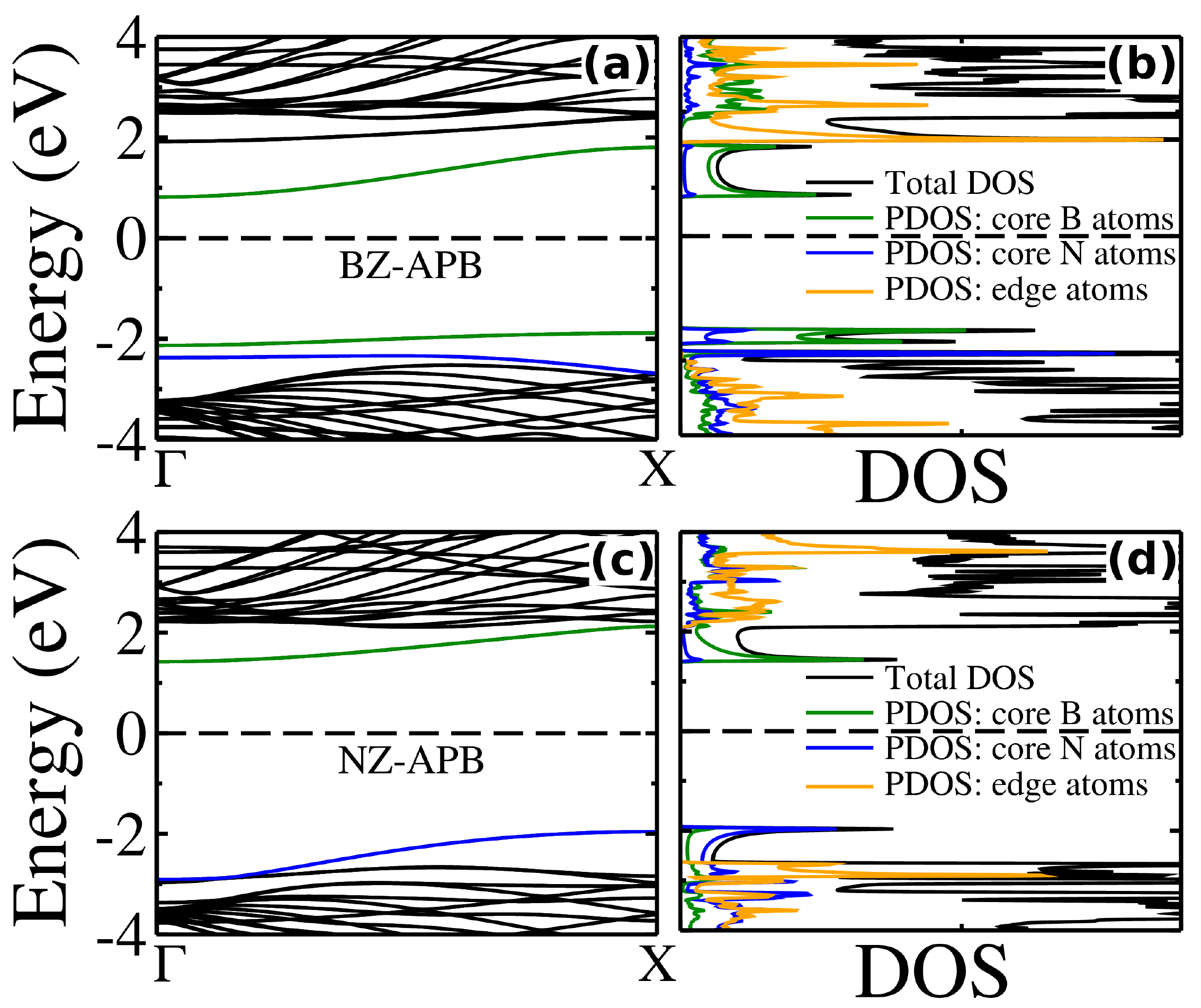}
\caption{Band structure along the $\Gamma$-X line (parallel to the APB
direction) and density of states: (a) and (b) for the BZ-APB; and (c)
and (d) for the NZ-APB. The contribution of the core-atom orbitals to
the total DOS is shown by blue (N orbitals) and green (B orbitals)
PDOS curves in (b) and (d). Defect bands in the gap are shown by green
and blue curves, according to the dominant atomic-orbital contribution
in each case. The DOS features associated to the ribbon-edge states
are shown by orange curves in (b) and (d).}
\label{fig7}
\end{figure}

The NZ-APB and BZ-APB display much richer electronic structures. The
bands and DOS curves for these boundaries are shown in
Fig.~\ref{fig7}. We include in this figure the electronic states of the
boundaries obtained from the ribbon calculations, from which we can
identify the states from each of the boundaries individually. One
observation regarding the consistency of this procedure is that the
electronic states associated with the edges of the ribbon are either
shallow or resonant with the bulk bands, and do not mix with the
defect bands, as can be observed in the PDOS plots in
Fig.~\ref{fig7}(b) and (d) for the NZ-APB and the BZ-APB,
respectively. This allows us to easily identify the gap states
associated with the NZ-APB and BZ-APB. Moreover, we verify that the
electronic structure we obtain from the supercell calculation [not
shown in Fig.~\ref{fig7} for conciseness], which contains the states
from both boundaries, is very consistent with a superposition of the
corresponding states from the ribbons for each boundary, in the range
of energies of the defect gap states shown in this figure.

Generally, we can see in Fig.~\ref{fig7} that the NZ-APB and BZ-APB
boundaries introduce much deeper defect bands into the gap of the BN
bulk, resulting in a much larger reduction of the electronic band gap
in the spatial region surrounding either the NZ-APB or the
BZ-APB~\cite{xli} than in the case of the A-APB. Furthermore, the
dispersions for these bands are much larger, indicating carrier with
potentially larger mobilities than in the A-GB states.

The electronic structure of the BZ-APB ribbon is shown in
Fig.~\ref{fig7}(a) and (b). Because boron-boron bonds tend to be much
longer than the other bonds in these BN systems, the BZ-APB introduces
a large compressive strain in its neighborhood, and we observe three
defect bands connected with this boundary. Starting from the lower
part of the gap, we observe a shallower band with an extended van Hove
singularity connected to a large flat portion of this band, starting at
the $\Gamma$ point and extending up to the k-point at $\sim$0.6 of the
$\Gamma$-X line, after which it disperses down and mixes with the bulk
states, when reaching the X point at the edge of the one-dimensional
BZ. This band derives primarily from the orbitals of the N atoms
located on the BN zigzag chains in the core of the BZ-APB ($\sim$58\%
of the total DOS), with smaller contributions from the zigzag B atoms
($\sim$8\%), and from the two atoms forming the B$_2$ dimer at the
geometric center of the defect ($\sim$3\%).

The BZ-APB also introduces a deeper flat band lying $\sim$0.5 eV above
the VBM, with a total dispersion of $\sim$0.2 eV. This band is very
strongly localized on the B$_2$ dimer at the center, with $\sim$75\%
of the total DOS deriving from the orbitals of the dimer atoms, with
smaller contributions of $\sim$12\% and $\sim$4\% from the N and B
atoms on the BN zigzag chains in the core, respectively. The
characteristic 1D van Hove singularities associated with the minimum
and maximum of this band can be observed in the DOS plot in
Fig.~\ref{fig7}(b).

In the upper part of the gap, the BZ-APB introduces a deep dispersive
band that lies $\sim$1.7 eV below the CBM, at the $\Gamma$ point, and
at the X point it reaches its maximum of $\sim$0.7 below the CBM, for
a total dispersion of 1 eV. This band is composed mostly of orbitals
of the B atoms at the BZ-APB core, with 53\% of the DOS coming from
the B$_2$ dimer at the center and 25\% from the B atoms on the BN
zigzag chains shouldering the B$_2$ dimer. A smaller contribution of
8\% derives from the N core atoms on the zigzag chains.

Electronic bands, DOS, and PDOS for the NZ-APB ribbon are shown in
Fig.~\ref{fig7}(c) and (d). In the lower part of the gap this
boundary introduces a band with a maximum energy of $\sim$0.7 eV above
the VBM, near the edge of the BZ at the X point, that shows a
dispersion of $\sim$1.0 eV. Its minimum is at the $\Gamma$ point where
it becomes resonant and mixed with the bulk states in the VBM
region. For most of the $\Gamma$-X line of the 1D BZ this band is deep
in the gap and strongly localized, with $\sim$88\% of the DOS
concentrated on the NZ-APB core atoms ($\sim$73\% on the N core
atoms). In the DOS plot, this bands shows a 1D van Hove singularity
above the peak corresponding to the VBM.

In the upper part of gap, the NZ-APB also introduces a band near the
CBM that shows similar characteristics as the above one, being deep in
the gap near $\Gamma$ and becoming shallow and mixed with the bulk
states when it reaches the edge of the BZ at the X point. This band is
very strongly localized on the B and N atoms along the zigzag chain in
the core of the NZ-APB, with $\sim$84\% of the DOS concentrated on the
orbitals of the B atoms and $\sim$7\% on the orbitals of the N
atoms. The corresponding 1D van Hove singularity is seen in the DOS
plot in Fig.~\ref{fig7}(d).
\begin{figure}
\vspace{0.1cm}
\centering \includegraphics[width=8cm]{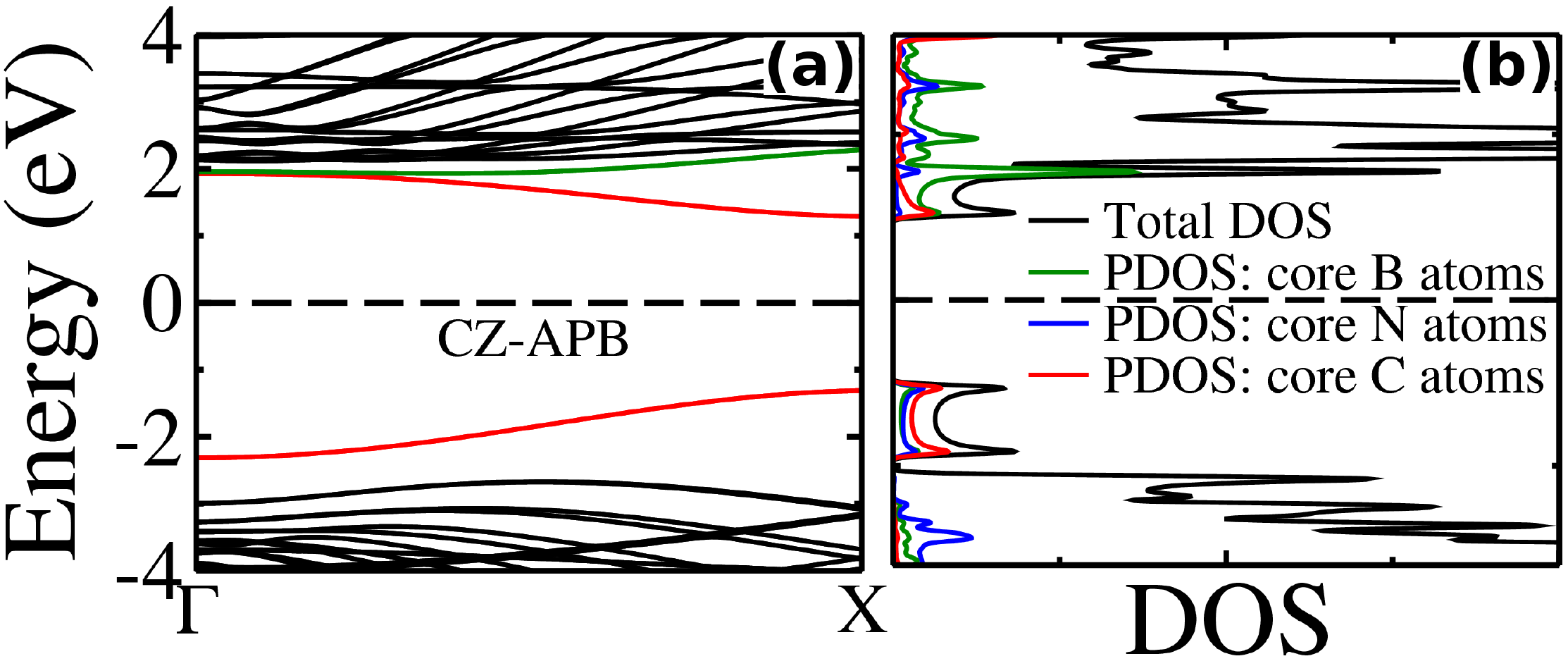}
\caption{Band structure along the $\Gamma$-X line (parallel to the APB
direction) and density of states for the CZ-APB. In (b) the
contribution of the core-atom orbitals to the total DOS is shown by
blue (N orbitals), green (B orbitals), and red (C orbitals) PDOS
curves. Defect bands in the gap are shown by green and red curves,
according to the dominant atomic-orbital contribution in each case.}
\label{fig8}
\end{figure}

The electronic states of the CZ-APB show a similar pattern. We observe
the appearance of two bands that are reminiscent of those of a
one-dimensional dimerized chain of carbon atoms, with a gap of
$\sim$2.6 eV at the BZ edged, and dispersions of $\sim$1.0~eV (0.6~eV)
for the lower (higher) band.  The higher band is derived equally from
C and B orbitals, while the lower band is dominated by the C orbitals,
but with sizeable contributions from B and N orbitals.  Both bands are
strongly localized on the defect core, with 78\% and 88\% of the total
DOS deriving from the core-atom orbitals.

In conclusion, our calculations indicate that the relative stability
of antiphase boundaries (APB) with armchair and zigzag chiralities in
planar boron nitride (BN) is determined by B and N chemical potential
conditions (emulating the corresponding synthesis conditions). In an
N-rich synthesis environment, a zigzag APB with pentagonal and
octagonal rings in its N-rich core is the most stable structure in
this material (among the models we considered), while for B-rich or
intrinsic conditions a stoichiometric armchair APB with tetragons and
octagons in its cores is the most stable. Such stability transition as
a function of B and N chemical potentials is shown to arise from a
competition between ``wrong-bond'' (homopolar B-B and N-N bonds) and
elastic-energy costs at the core of the APBs. This is contrasted with
analogous cases of extended defects in graphene, where the geometry
with pentagonal and octagonal rings is the most stable. In the
presence of carbon dopants, a carbon-doped zigzag APB becomes the most
stable boundary for the last third part of the interval of realistic B
and N chemical potentials values, comprising the last part of the
intrinsic region plus the N-rich region. The electronic structure of
the APBs in monolayer BN is shown to be particularly distinct, with
the pentagonal-octagonal APB depicting a defect-like electronic
structure with deep bands in the band gap, and the
tetragonal-octagonal APB depicting bulk-like shallow electronic bands
derived from the corresponding band edges.

{\it Note added: A brief account of the present work was previously
presented in a scientific meeting, and a summary was published in
Ref.~\onlinecite{lidia-aps}. During the preparation of this
manuscript, we became aware of a recent work~\cite{yliu} that also
considers the A-APB in BN, of which we had no previous knowledge.}

\begin{acknowledgments}
The authors acknowledge support from the Brazilian agencies CNPq,
FAPEMIG, Rede de Pesquisa em Nanotubos de Carbono, INCT de
Nanomateriais de Carbono, and Instituto do Mil\^enio em
Nanotecnologia-MCT.
\end{acknowledgments}

\end{document}